\documentstyle[preprint,aps]{revtex}

\tightenlines

\begin{document}

\begin{center}
\begin{large}
{\bf Orbital order and ferrimagnetic properties of the new
compound Sr$_8$CaRe$_3$Cu$_4$O$_{24}$}
\end{large}


Xiangang Wan, Masanori Kohno and Xiao Hu


Computational Materials Science Center, National Institute for
Materials Science, Tsukuba 305-0047, Japan

\end{center}

\begin{abstract}
By means of the LSDA+U method and the Green function method, we
investigate the electronic and magnetic properties of the new
material of Sr$_8$CaRe$_3$Cu$_4$O$_{24}$. Our LSDA+U
calculation shows that this system is an insulator with a net
magnetic moment of 1.01 $\mu_{\rm B}$/f.u., which is in good
agreement with the experiment. Magnetic moments are mainly located
at Cu atoms, and the magnetic moments of neighboring Cu sites
align anti-parallel. It is the non-magnetic Re atoms
that induce an orbital order of $d$ electrons of Cu atoms,
which is responsible for the strong exchange interaction and the
high magnetic transition temperature. Based on the LSDA+U results,
we introduce an effective model for the spin degrees of freedom,
and investigate the finite-temperature properties by the Green
function method. The obtained results are consistent with the
experimental results, indicating that the spin-alternating
Heisenberg model is suitable for this compound.
\end{abstract}

75.30.Et, 75.10.Jm, 75.10.-b, 71.70.Gm

\newpage

Perovskite transition-metal oxides have been investigated
intensively\cite{Perovskite}. The discoveries of high-$T_{\rm c}$
superconductivity\cite{superconductor} and colossal
magnetoresistance (CMR)\cite{GMR-1} in this family of materials
pose significant challenge to theoretical understanding of the
strongly correlated electron systems.
\par
In certain parameter regimes, strong Coulomb repulsions make
electrons localized and the materials behave as Mott insulators.
In this phase, low-energy excitations are dominated by the
superexchange interaction\cite{Anderson}, and the collective excitations
of spin degrees of freedom govern the physics. Undoped high-$T_{\rm c}$
superconductors (HTC) are such typical examples. Their properties
are well described by a spin-1/2 antiferromagnetic (AFM) Heisenberg
model on a square lattice. It is widely believed that properties
of the undoped systems are closely related to the origin of
high-$T_{\rm c}$ superconductivity\cite{ZhangRice}.
\par
It is also well known that the charge distribution in perovskite
transition-metal oxides is influenced strongly by the crystal
field, i.e. the degeneracy of the 3$d$ electrons is lifted by
lattice distortions of the perovskite structure in various ways.
Therefore, the spin, charge and orbital degrees of freedom are
intervened with each other, which make the physics of the
perovskite transition-metal oxides very rich. These features are
shared by materials in which CMR are observed\cite{GMR-2}.
\par
As will be revealed in this Letter, the recently synthesized
material Sr$_8$CaRe$_3$Cu$_4$O$_{24}$, which forms the cubic
perovskite structure as shown in Fig. 1(a), is another member of
this family of materials\cite{Takayama-Muromachi}. The new
material is similar to the undoped HTC in the following aspects:
(1) perovskite structure, (2) magnetic insulator, (3) Cu
surrounded by O. However, it is different in the points that this
material is three-dimensional and has nonzero magnetization in the
ground state $M\simeq 0.95 \mu_{\rm B}$ per formula unit (f.u.)
Also, this compound is peculiar since ferromagnetic (FM) cuprates
are very rare, and has an unusually high magnetic transition
temperature $T_{\rm c}\simeq 440$ K. Curie temperatures $T_{\rm
c}$ of known FM cuprates are usually quite low.  For example, the
$T_{\rm c}$ of La$_4$BaCu$_2$O$_{10}$, K$_2$CuF$_4$ and SeCuO$_3$
are 5, 6.5, and 26 K, respectively\cite{Tc}. Therefore, it is
interesting to see why a strong FM state is realized in this
material. Although the response of the compound to carrier doping
is also very interesting, we will concentrate here on the undoped
material. It is shown that both the electron-electron correlation
and the coupling between orbital and spin degrees of freedom are
important for this compound, which results in an spin-alternative
Heisenberg Hamiltonian.

\par

We calculate the electronic and magnetic structures of
Sr$_8$CaRe$_3$Cu$_4$O$_{24}$ by using the WIEN2K
package\cite{Blaha}, which is an implementation of the
density-functional APW+lo method\cite{Singh}. The MT sphere radii
of 2.0, 2.0, 2.0, 1.9 and 1.5 a.u. are chosen for the Ca, Cu, Sr,
Re, and O atoms, respectively, with $RK_{\rm max}$=7.0, which
results in about 5,200 LAPWs and local orbitals per cell. We use
700 $k$ points in the Brillouin zone. As for the
exchange-correlation potential, we adopt the standard generalized
gradient approximation (GGA)\cite{Perdew}. Our GGA calculation
predicts this material as a metal inconsistent with the
experiment, which indicates that the electron-electron
interactions are important for this compound. Thus, we use the
LSDA+U method\cite{Anisimov} to take them into account, and adopt
$U$=10 eV and $J$=1.20 eV for the $d$ orbital of Cu\cite{U}.
\par
In order to investigate the magnetic structures and interactions
of Sr$_8$CaRe$_3$Cu$_4$O$_{24}$, we calculate both AFM and FM
configurations for Cu1 and Cu2 moments. The results are shown in
Table I. Magnetic moments are mainly located at Cu1 and Cu2 sites
in both configurations with the magnitude almost independent of
the configuration. The magnetic moments at Ca, Sr, O1 and O3 are
smaller than 0.001 $\mu_{\rm B}$. Thus, they are negligible
compared to the moments at Cu. O2 carries a small but nonvanishing
moment due to large hybridization with Cu1 and Cu2. Our numerical
results show that the AFM configuration is the ground state with
net magnetic moment $M$=1.01 $\mu_{\rm B}$/f.u. This is
quantitatively consistent with the experimental one. Also, the
ground state is an insulator with the energy gap of 1.68 eV
consistently with the experimental
result\cite{Takayama-Muromachi}. It should be noted that the
magnetic moment at Re atom is very small (less than 0.004
$\mu_{\rm B}$) in both configurations. This feature is contrasted
with the ordered double perovskite A$_2$Fe$M$O$_6$, where due to
the hybridization with $d$ orbitals of Fe, the non-magnetic ion
$M$ possesses a large magnetic moment and plays a significant role
in the magnetic properties\cite{Sarma,Fang}. We also performed
calculations for $U$=8 and 12 eV to cover a wide range of
generally accepted values of correlations. Numerical results show
that the magnetic moments are not sensitive to $U$. With
decreasing $U$, the energy difference between AFM and FM ($\Delta
E$) increases slightly. For example, the calculated $\Delta E$ for
$U$=8 eV is higher than that of $U$=10 eV by about 20$\%$.
\par
Figure 2(a) shows the calculated charge density plotted in (010)
plane of Fig. 1(a). The charge density along the Re-O3 bonds is
larger than that along Cu2-O3, indicating a stronger bond for
Re-O3 than that for Cu2-O3. Therefore, the O3 atom is attracted
toward Re. This induces a Jahn-Teller distortion in the oxygen
octahedron centered at Cu2 consisting of four O3 and two O2 atoms
with the bond length of Cu2-O2 smaller than that of Cu2-O3.
Consequently, the $e_g$ orbital of Cu2 splits into
$d_{3z^{2}-r^{2}}$ and $d_{x^{2}-y^{2}}$. The former which points
to O2 has a higher energy than the latter. Thus as shown in Fig.
3(b), the partially occupied orbital in Cu2 is the minority spin
(spin-up) $d_{3z^2-r^2}$, and the charge density distribution of
Cu2 is anisotropic as shown in Fig. 2(a). In contrast, three
crystallographic directions of Cu1 are completely equivalent, and
the oxygen octahedron centered at Cu1 site is free of any
distortion. Therefore, the $e_g$ orbital of Cu1 is still fully
degenerate, and all $d$ orbitals except for the minority spin
(spin-down) in the $e_g$ orbital are fully occupied as shown in
Fig. 3(a). In addition, the 2{\it p} states of O2 distribute
mainly in the energy range from -7.0 to 0.0 eV, and are almost
full-filling.

\par
The spin density shown in Fig. 2(b) clearly indicates that the
magnetic moments of the Cu1 and Cu2 are carried mostly by $e_g$
and $d_{3z^2-r^2}$ orbitals, respectively. In Fig. 4, we
schematically show the partially occupied $d$ orbitals of Cu1 and
Cu2 and the almost fully occupied $p_z$ orbital of O2 for the
material. It is interesting to observe that an orbital ordering
appears at Cu sites. Namely, the minority spin $d_{3z^2-r^2}$
orbital of Cu2 and the $e_g$ orbital of Cu1 are pointing to the O2
as shown in Fig. 4. Both of them are less than half-filling.
Meanwhile, the fully occupied $p_z$ orbital of O2 points to the
neighboring Cu1 and Cu2 sites. Therefore, the spin-up and -down
$p_z$ orbitals of O2 strongly overlap with the spin-up
$d_{3z^2-r^2}$ of Cu2 and the spin-down $e_g$ orbital of Cu1,
respectively, to form a rather strong $pd\sigma$ hybridization.
This results in a strong exchange interaction between the magnetic
moments at Cu1 and Cu2 and a high transition temperature.
\par
In order to further clarify the effects of Re, we perform a
calculation for an artificial structure, where the Jahn-Teller
distortion in oxygen octahedron centered at Cu2 is removed. We
find that the orbital ordering still survives, and the results are
very similar to that of the real structure. We then change Re to
Technetium, which is the element at the same column in the
periodic table. Our calculation shows that the partially occupied
orbital of Cu2 changes from $d_{3z^2-r^2}$ to $d_{x^2-y^2}$.
Consequently, the spin and the orbital are almost decoupled, and
the spin interaction becomes much weaker. As a result, the energy
difference between AFM and FM configurations is reduced greatly.
Therefore, the Re atoms play an important role for the orbital
ordering to cause the unusually high $T_{\rm c}$, although their
magnetic moments are very small.
\par
Since the magnetic moments are almost localized at Cu sites, the
effective model for the spin degrees of freedom of this material
is expected to be a Heisenberg model, where neighboring localized
spins at Cu sites interact antiferromagnetically. Although, the
simple ionic model (Cu$^{2+}$-O$^{2-}$ or Cu$^{3+}$-O$^{2-}$) is
not exact due to the large hybridization between Cu and O, we
still take spins at Cu sites as a good quantum number similar to
the case of the undoped HTC. We consider the following two cases:
(1)$S_1$=1, $S_2$=1/2 and (2) $S_1$=$S_2$=1/2, where $S_i$ denotes
the spin length at Cu$i$ site. The magnetization in the unit cell
for the AFM state and that of the FM state are obtained as (1)
$M_{\rm AFM}$=1/2, $M_{\rm FM}$=5/2 and (2) $M_{\rm AFM}$=1,
$M_{\rm FM}$=2. The LSDA+U result is $M_{\rm AFM}$=1.01 $\mu_{\rm
B}\simeq$0.5, $M_{\rm FM}$=5.01 $\mu_{\rm B}\simeq$2.5 as shown in
Table I. Hence, case (1) is reasonable. Thus, the effective
Hamiltonian becomes the following:
\begin{equation}
H=J_{\rm eff}\sum_i{ S}_i\cdot\sum_p s_{i+\frac{p}{2}},
\end{equation}
where $J_{\rm eff}$ is the effective exchange interaction, $i$
runs over Cu1 sites, and $p$ denotes the unit vectors of the unit
cell in the lattice shown in Fig. 1(b) ($p$=$\pm{\hat x}, \pm{\hat
y}, \pm{\hat z}$). Here, $S_i$ and $s_{i+\frac{p}{2}}$ denote the
spin operator of $S$=1 at Cu1-site($i$) and that of $S$=1/2 at
Cu2-site($i$+$\frac{p}{2}$), respectively.
\par
Since this model is on a bipartite lattice and belongs to the
family in which the Marshall-Lieb-Mattis theorem
holds\cite{Marshall}, the ground state of this model
is proven to have spin $S$=1/2(=1/2$\times$3-1) in any size of
systems. This results in spontaneous magnetization $M$=1 $\mu_{\rm B}$
in the bulk limit, and well explains the experimental result
($M \simeq 0.95 \mu_{\rm B}$\cite{Takayama-Muromachi}).
\par
Since the difference of the magnitude of magnetic moments at Cu
sites between FM and AFM states is small as shown in Table I, the
energy difference between FM and AFM $\Delta E$=0.036 Ry=5,679 K
can be assigned to the exchange interaction. By diagonalizing the
Hamiltonian (1) in the unit cell, we evaluate the energy
difference between FM and AFM configuration as $8J_{\rm eff}$.
Hence, we estimate $J_{\rm eff}$=$\Delta E$/8=710 K. Using the
above effective Hamiltonian, we investigate magnetic properties of
Sr$_8$CaRe$_3$Cu$_4$O$_{24}$. In order to take quantum
fluctuations into account, we apply the Green function
method\cite{GF}. In this method, we construct sixteen Green
functions of Cu1-spins and Cu2-spins. We use Tyablikov's
decoupling\cite{Tyablikov} to obtain closed coupled equations for
Green functions. Following Callen's scheme\cite{Callen}, we obtain
$\langle s^z\rangle$ and $\langle S^z\rangle$ as a function of
spin-wave dispersion relations $E_k$. Here, $E_k$ is also a
function of $\langle s^z\rangle$ and $\langle S^z\rangle$. Thus,
we solve these coupled equations numerically to obtain $\langle
s^z\rangle$ and $\langle S^z\rangle$ at finite temperatures.
\par
The result of the Green function method for the spontaneous
magnetization is shown in Fig. 5. In the low-temperature limit,
spontaneous magnetization $M$ obtained by the Green function
method is about $1\mu_{\rm B}$, which is consistent with the exact
result $M$=1 $\mu_{\rm B}$ and the experimental result
$M\simeq$0.95 $\mu_{\rm B}$. Magnetic transition temperature
$T_{\rm c}$ is calculated as $T_{\rm c}=448$ K by the Green
function method, which is close to the experimental result $T_{\rm
c}\simeq$440 K. Thus, the magnetic properties known so far in
Sr$_8$CaRe$_3$Cu$_4$O$_{24}$ are explained by the spin-alternating
Heisenberg model defined in eq.(1).
\par
Here, we summarize the mechanism of ferrimagnetism and the high magnetic
transition temperature $T_{\rm c}$ for Sr$_8$CaRe$_3$Cu$_4$O$_{24}$.
The mechanism of the AFM coupling is the superexchange\cite{Anderson}.
The large exchange constant $J_{\rm eff}$ is due to the large overlap
between Cu and O2 orbitals, and to the localization of Cu-spins
by strong correlations. This situation is similar to the case of the
undoped HTC. In contrast to the undoped HTC, the number of sublattice
sites is different. This causes ferrimagnetism. Imbalance of the spin
length between Cu1 and Cu2 sites would also be a reason for stabilization
of the ferrimagnetic state.
\par
In summary, using the LSDA+U method and the Green function method,
we have investigated the electronic and magnetic properties of
Sr$_8$CaRe$_3$Cu$_4$O$_{24}$. Our results show that magnetic moments are
almost located at Cu sites, and the moments of Cu1 and Cu2 align
anti-parallel due to the superexchange interaction\cite{Anderson}.
The ground state is a ferrimagnetic insulator with a net magnetic moment
of 1.01 $\mu_{\rm B}$/f.u., which is in good agreement with the
experimental results. We find that the non-magnetic transition
metal Re plays an important role in determining the magnetic properties
of this material. Namely, Re induces orbital orderings of the $d$ orbitals
in Cu. This results in the strong exchange coupling between the
magnetic moments of Cu and the unusually high $T_{\rm c}$. Based on the
LSDA+U results, we have introduced an effective model for the spin degree
of freedom, and investigated the magnetic properties by the Green function
method. The ground-state magnetization of this model is shown to be exactly
1 $\mu_{\rm B}$, which is consist with the experimental result. The obtained
$T_{\rm c}$ is also consistent with the experimental one. Therefore,
the spin-alternating Heisenberg model is suitable for describing the magnetic
properties of this compound.

We acknowledge Dr. E. Takayama-Muromachi for bringing our
attentions to this material and for valuable discussions.
Dr. M. Arai and Dr. M. Katakiri are appreciated for technical helps.
This work was partially supported by Japan Society for the Promotion of
Science (Grant-in-Aid for Scientific Research (C) No. 15540355).

\bibliography{apssamp}

\begin{table}
\caption{Calculated total energy, E$_{tot}$ relative to the energy
of AFM configuration in units of Ry, the total magnetic moments
per unit cell $\mu_{tot}$, the magnetic moment at Cu1, Cu2 and O2
in units of $\mu_{B}$.}

\vskip5mm
\begin{center}
\begin{tabular}{lccccc}
              &  E$_{tot}$  & $\mu_{tot}$  & Cu1 & Cu2 &O2 \\
\hline
FM    & 0.036    & 5.01       &  1.15 &  0.84     &  0.14\\
AFM   & 0.0      &-1.01       &  1.09 & -0.81     &  0.07\\
\end{tabular}
\end{center}

\end{table}

\newpage
\begin{figure}
\caption{Crystal structure of Sr$_{8}$CaRe$_{3}$Cu$_{4}$O$_{24}$.
(a) Unit cell. (b) Unit cell of the effective model. }
\end{figure}

\begin{figure}
\caption{Contours for charge density in the (010) plane with
interval 0.03 e/bohr$^3$. (a) Total charge density. (b) Spin
density (spin-up charge density minus spin-down charge density).
The dotted lines are negative contours.}
\end{figure}

\begin{figure}
\caption{Projected density of state with Fermi energy at zero. (a)
For minority spin (spin-down) {\it d} orbital of Cu1. (b) For
minority spin (spin-up) {\it d} orbital of Cu2. }
\end{figure}

\begin{figure}\caption{
Schematic picture for the spin and orbital orders of
Sr$_{8}$CaRe$_{3}$Cu$_{4}$O$_{24}$ deduced from the LSDA+U
calculation. The gray symbols denote partially occupied {\it d}
orbital, and the dark ones denote full-occupied {\it p} orbital.
The arrows denote the directions of magnetic moments on Cu sites.}
\end{figure}

\begin{figure}
\caption{Temperature dependence of the  spontaneous magnetization.
Solid curve denotes the Green function result. Squares and the
dotted line are experimental results in Ref.[7].}
\end{figure}

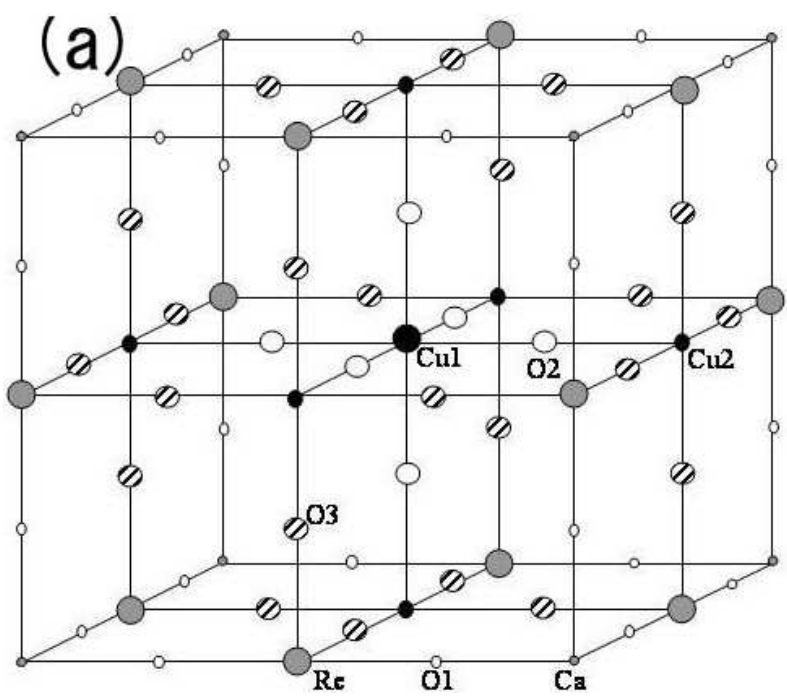
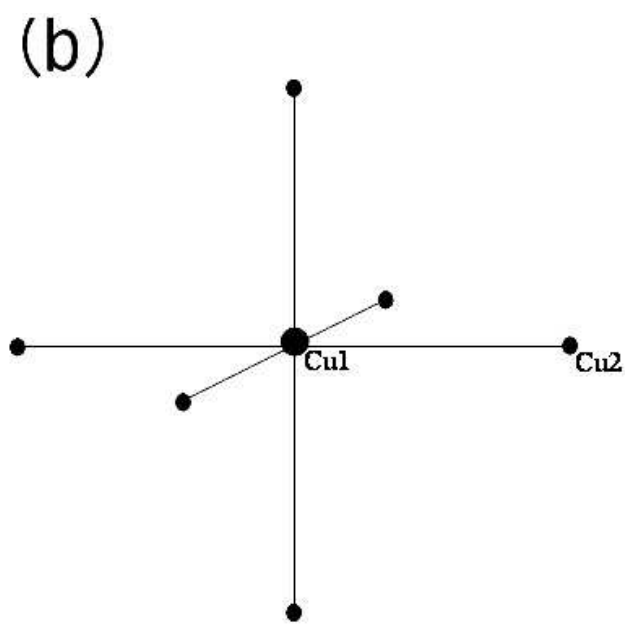
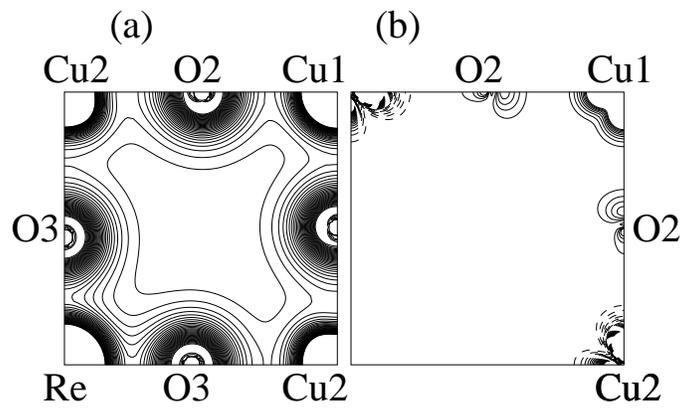
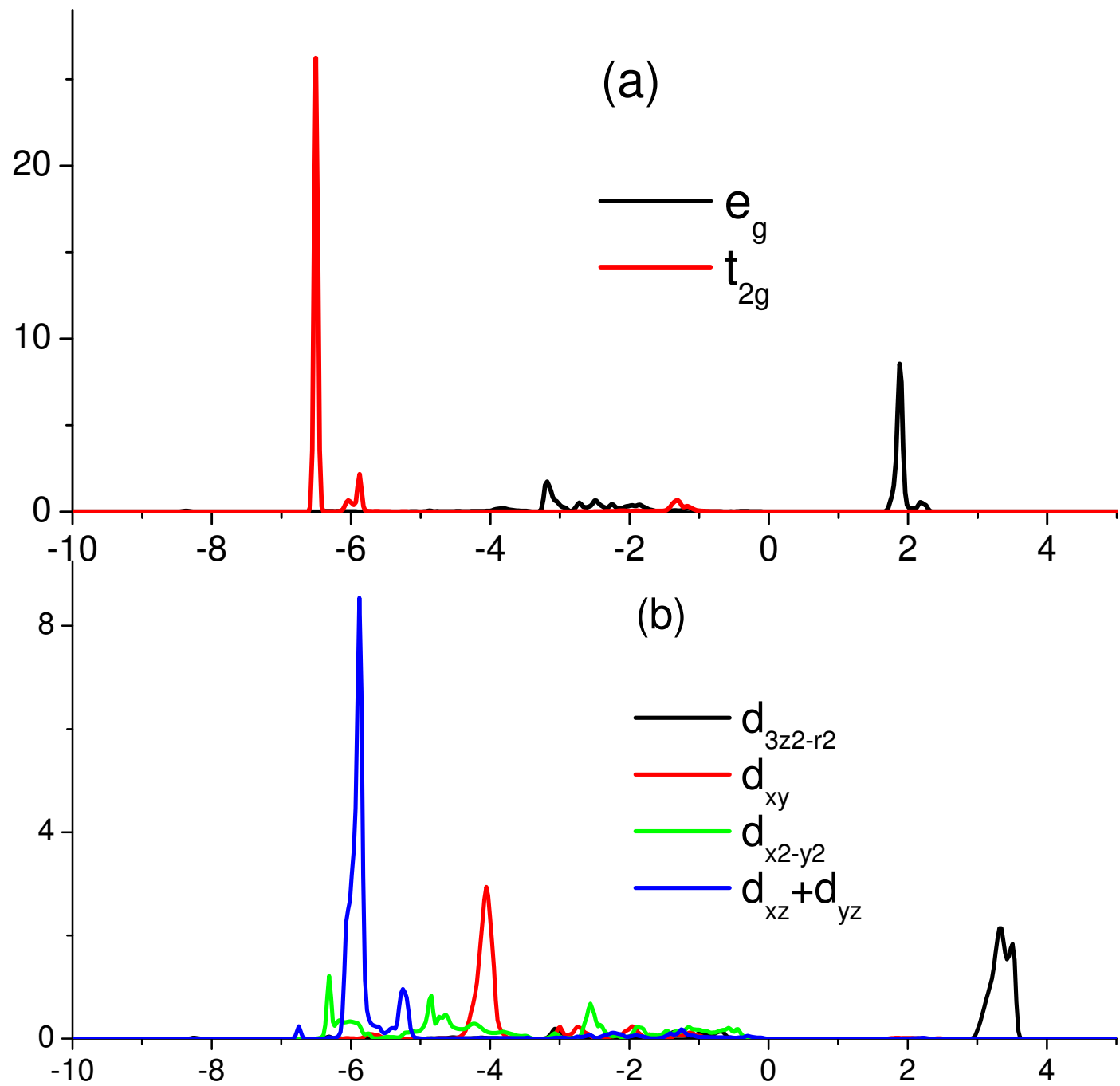
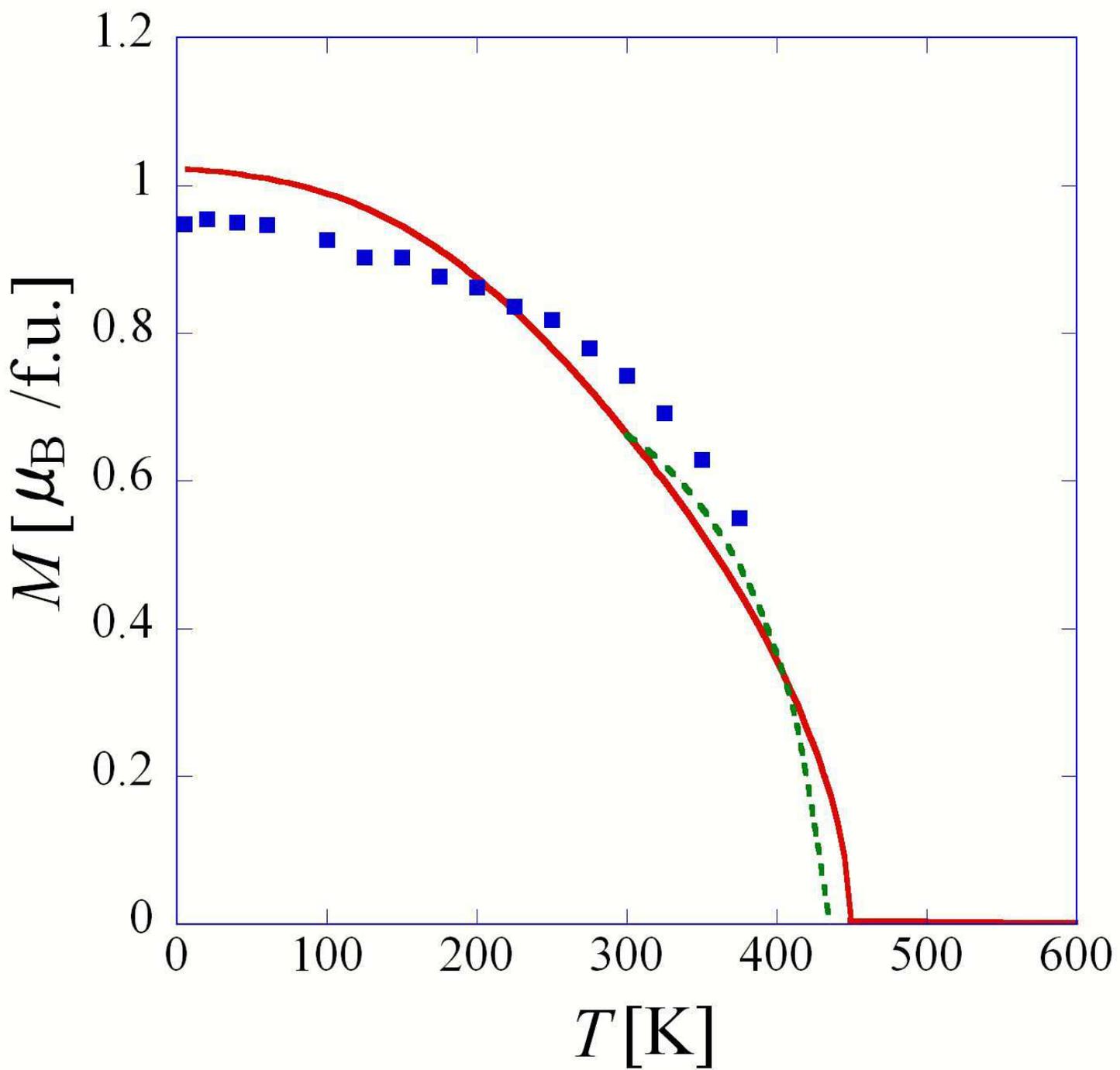
\end{document}